\documentclass[reprint,amsmath,amssymb,aps,prl,superscriptaddress]{revtex4-1}

\usepackage{times,color,amsmath}
\usepackage{subfigure,epsfig,graphicx,epstopdf}
\usepackage[colorlinks]{hyperref}
\hypersetup{linkcolor = {blue},	citecolor = {blue},	urlcolor = {blue}}
\usepackage{textcomp}
\usepackage{verbatim}

\newcommand{\affa}{\affiliation{Department of Applied Physics, Royal Institute of Technology, Albanova University Centre,
Roslagstullsbacken 21, 106 91 Stockholm, Sweden}}
\newcommand{\affb}{\affiliation{Department of Physics, Zhejiang Normal University, Jinhua 321004, China}}

\begin{document}

\title{Bloch--Siegert Physics in a Reconfigurable Photonic Binary Lattice}

\author{Ze-Sheng Xu} \email{zesheng@kth.se}\affa        
\author{Liwei Duan} \email{duanlw@gmail.com}\affb         \author{Rohan Yadgirkar} \affa 		 
\author{Andrea Cataldo} \affa    \author{Adrian Iovan} \affa
\author{Jun Gao} \affa
\author{Ali W. Elshaari} \email{elshaari@kth.se}  \affa

\preprint{APS/123-QED}

\date{\today}

\begin{abstract}
The Bloch–Siegert shift, a hallmark correction arising from counter-rotating interactions in driven two-level systems, has an exact counterpart in binary lattices under static forcing, where it governs resonant long-range tunneling between sites separated by odd lattice spacings. Here we report the first experimental realization of this correspondence using a 12-mode programmable photonic integrated circuit. By implementing a reconfigurable binary lattice with sub-percent control of on-site detuning, we observe coherent periodic jumps across four resonance orders and quantitatively verify the predicted period law over the full parameter space. The measured dynamics exhibit the extreme resonance sensitivity characteristic of Bloch–Siegert physics and agree closely with the level-anticrossing picture of the semiclassical Rabi model. Exploiting the underlying parity structure, we further convert intrinsically bidirectional oscillations into cascaded unidirectional transport through adaptive sign reversal of the staggered potential, achieving fidelities exceeding 0.95 and 0.98 on the same hardware platform. Our results establish programmable photonic lattices as a scalable testbed for strongly driven quantum-optical phenomena and Floquet-engineered transport.
\end{abstract}
\maketitle

\begin{figure*}[!t]
	\centering
	\includegraphics[width=0.95\linewidth]{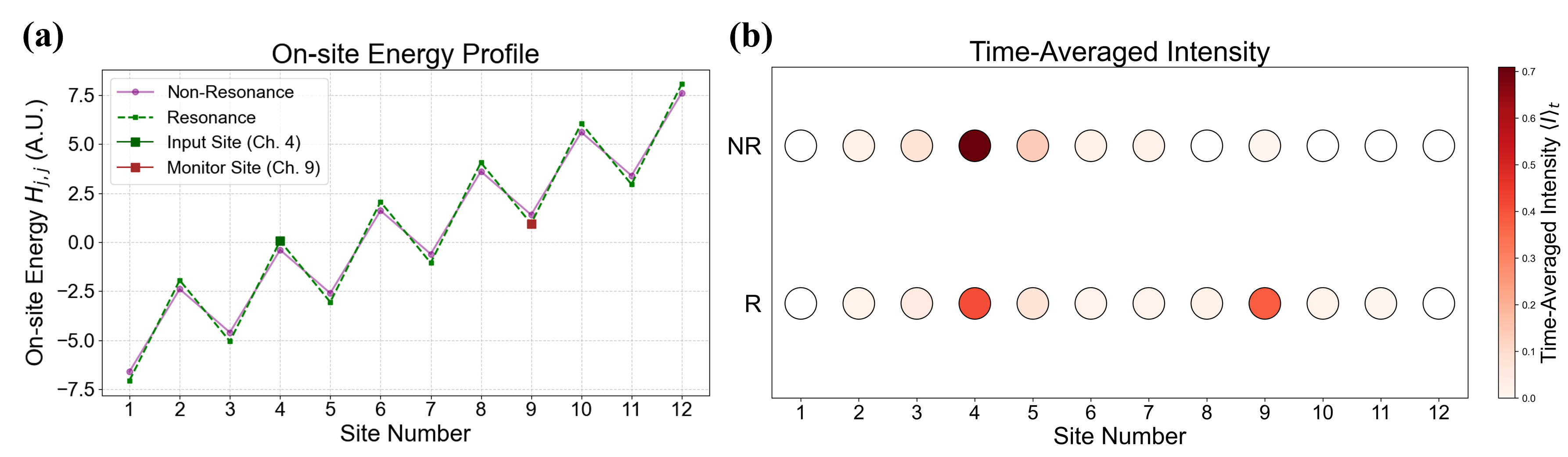}
	\caption{\textbf{Extreme sensitivity of resonant periodic jump to on-site energy detuning.}
{(a)~On-site energy profiles for both the resonant (R, green dashed) and non-resonant (NR, purple solid) conditions. Although the two configurations differ by only a marginal shift in the on-site energy mismatch $\epsilon$ --- well within what might be considered a small perturbation --- this minute detuning from the $m$-th order resonance condition $\epsilon \approx (2m+1)F$ precipitates a qualitative transition in the transport dynamics, as demonstrated in (b). The filled red and blue markers denote the input and monitor sites, respectively. (b)~The upper and lower panels show the time-resolved intensity distributions under the non-resonant and resonant conditions, respectively. Under non-resonance, the intensity remains predominantly localized at the input site with only weak spreading to neighboring sites. In sharp contrast, under resonance, the intensity undergoes coherent, long-range oscillations between site 4 and site 9, a hallmark signature of the $m=2$ periodic jump. The stark dichotomy between these two nearly identical on-site energy configurations highlights the sharp, resonance-like character of this transport phenomenon.}
}
	\label{f1}
\end{figure*}

\begin{figure*}[!t]
	\centering
	\includegraphics[width=0.8\linewidth]{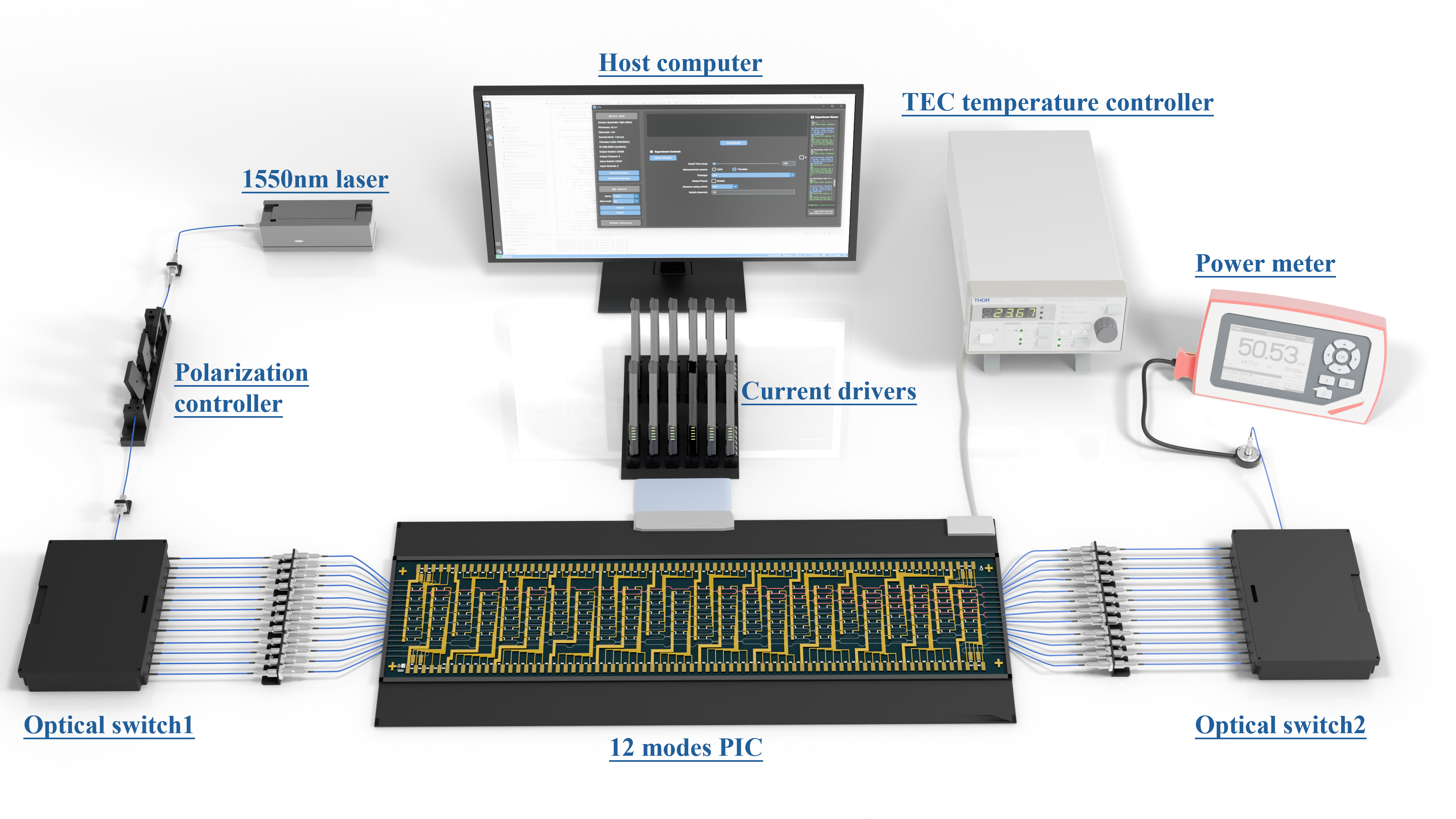}
	\caption{\textbf{Experimental setup. } 
		{An optical signal is generated by a short-pulse semiconductor source operating at $1550$ nm. This coherent light traverses a polarization controller, which is configured for the transverse electric (TE) mode, before being channeled into the input terminal of a $1 \times 12$ MEMS-based optical switch. This switch is utilized to selectively inject the signal into one of the twelve discrete input waveguides of the programmable photonic chip via edge couplers. The core element is the integrated photonic circuit, which employs a reconfigurable mesh of MZIs governed by high precision multichannel current drivers. The resultant output light from the chip is subsequently collected by a reciprocal $12 \times 1$ switch, also interfaced using edge couplers, and is directed from its common port to a high-sensitivity optical power meter for detection. System management and data acquisition are executed by a host computer (PC), which controls the switching elements, programs the thermal actuators to implement the target unitary operations, and logs the power meter measurements. Precise temperature stability is maintained by a solid-state thermoelectric cooler positioned beneath the chip, which is stabilized via a PID controller.}
	}
	\label{f2}
\end{figure*}

\begin{figure*}[!t]
	\centering
	\includegraphics[width=0.8\linewidth]{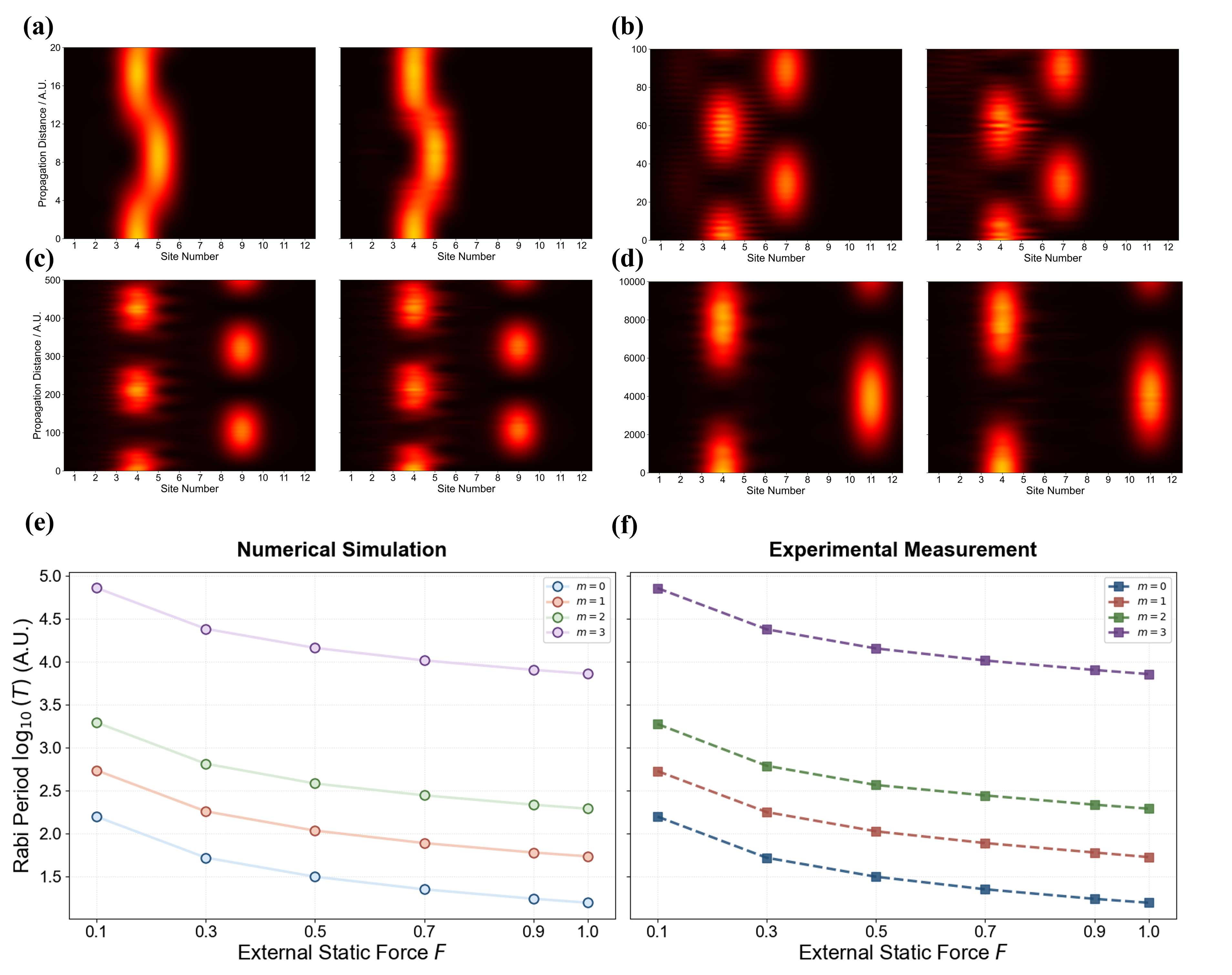}
	\caption{\textbf{Experimental observation of resonant periodic jumps across multiple orders and quantitative validation of the jump period.}
{(a)--(d)~Simulated (left) and experimentally measured (right) time-resolved intensity distributions for resonance orders $m = 0, 1, 2, 3$ at $F = 0.9$, with input fixed at site 4 and monitor sites at 5, 7, 9, and 11, respectively. (e)~Theoretically calculated jump periods $T = 2\pi/\Delta_{\min}$ as a function of static force $F$ for all four resonance orders. (f)~Corresponding experimentally fitted periods obtained via FFT followed by sinusoidal fitting. The near-perfect overlap between (e) and (f) across all orders and force values confirms the theoretical prediction with high quantitative accuracy, with deviations remaining well within experimental uncertainty.}
}
	\label{f3}
\end{figure*}

\begin{figure*}[!t]
	\centering
	\includegraphics[width=0.8\linewidth]{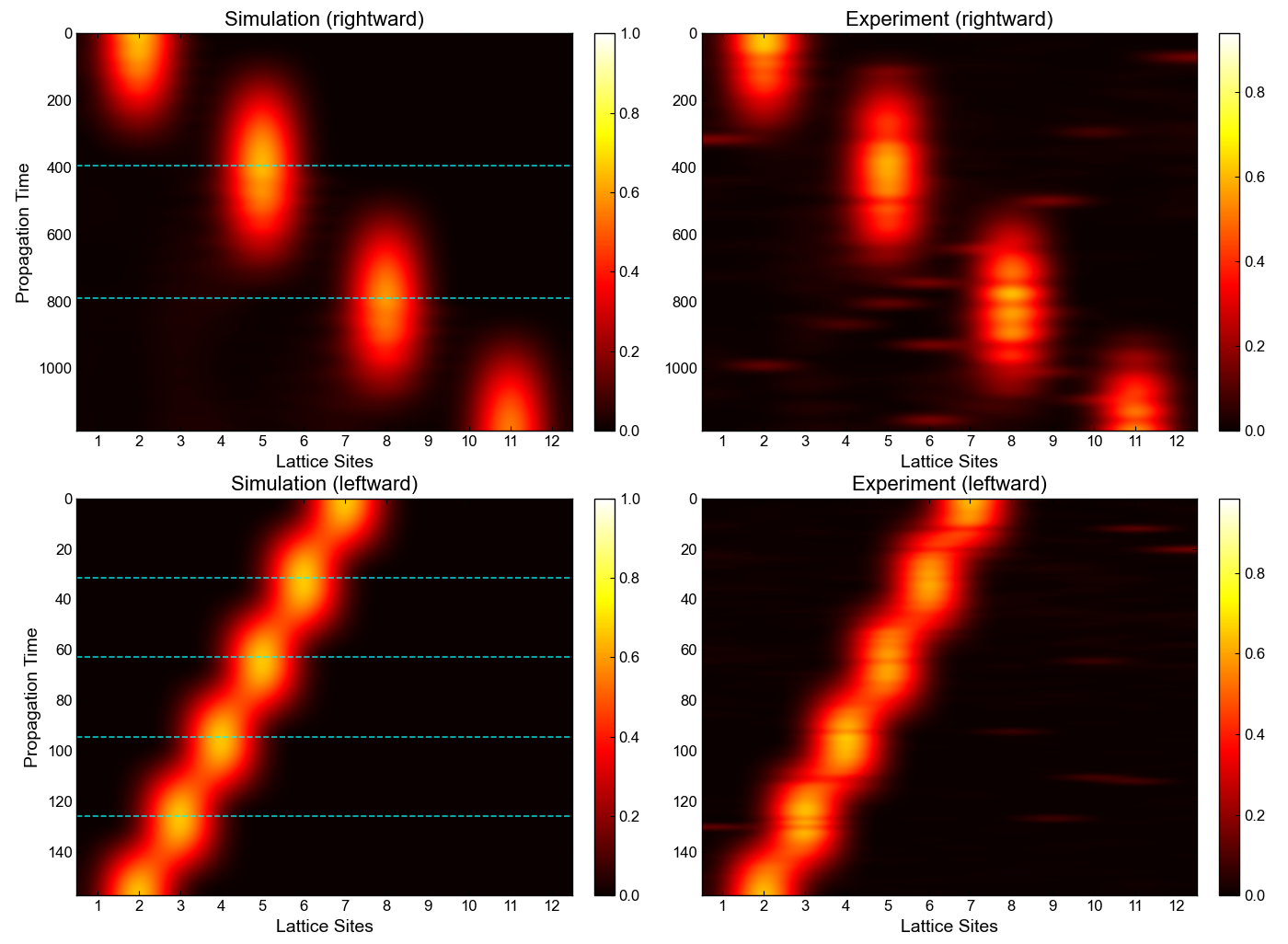}
	\caption{\textbf{Cascaded unidirectional transport via adaptive $\epsilon$ sign-flipping.}
		{Top row: first-order ($m=1$) rightward transport along the chain $2\!\to\!5\!\to\!8\!\to\!11$; bottom row: zeroth-order ($m=0$) leftward transport along $7\!\to\!6\!\to\!5\!\to\!4\!\to\!3\!\to\!2$. Left panels: numerical simulation; right panels: experimental intensity maps. Cyan dashed lines mark the per-stage adaptive flip times $t^*_k$ at which $\epsilon \to -\epsilon$. After each flip, the destination site acquires A-type sublattice character under the updated Hamiltonian, becoming the resonant entry for the next hop and suppressing the reverse path. The $m=1$ case uses $\epsilon/F \approx 2.905$, $V/F = 0.25$, with initial transfer fidelity exceeding $0.95$; the $m=0$ case uses $\epsilon/F \approx 1.0$, $V/F = 0.05$, with per-stage fidelity exceeding $0.98$. The two protocols differ in resonance order, propagation direction, and characteristic timescale ($t^* \approx 395$ vs.~$t^* \approx 31$ in units of $F^{-1}$), yet both run on the same 12-channel platform without hardware reconfiguration.}
	}
	\label{f4}
\end{figure*}

\begin{figure*}[!t]
	\centering
	\includegraphics[width=0.8\linewidth]{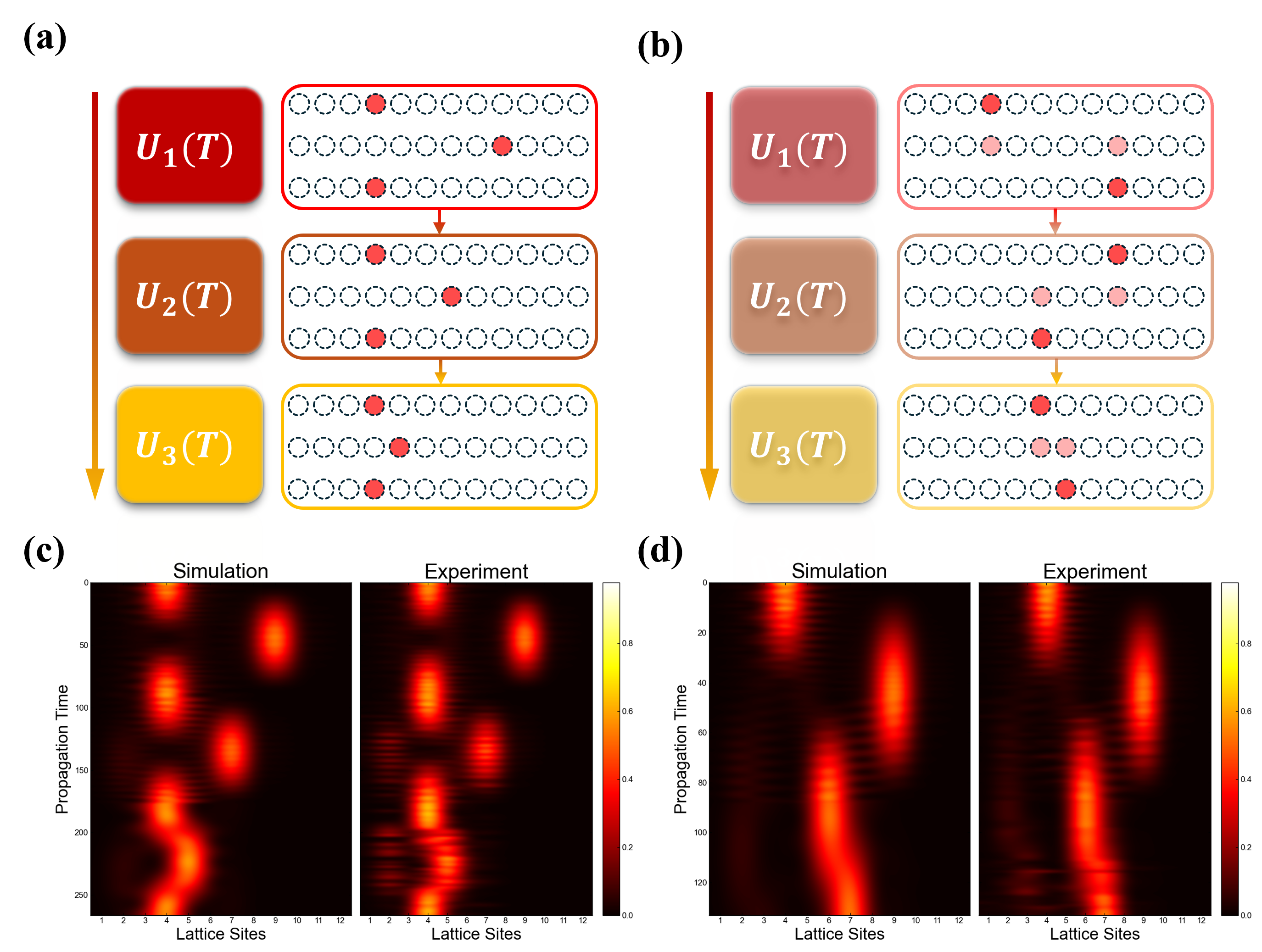}
	\caption{\textbf{Discrete Floquet evolution under full and half resonance periods.}
		{\textbf{(a)}~Sequential application of full-period unitary operators $U_m = \exp(-iH_m T_m)$; each block contains 50 discrete evolution steps, with the right panel detailing the intra-block dynamics. \textbf{(b)}~Half-period protocol ($r = 0.5$), in which the field advances progressively toward the lattice centre at each step rather than returning to the input. \textbf{(c),(d)}~Numerically simulated (upper) and experimentally measured (lower) intensity distributions for the full-period and half-period protocols, respectively. Excellent agreement in both cases confirms high-fidelity unitary implementation on the PIC.}
	}
	\label{f5}
\end{figure*}

The Bloch--Siegert shift---the frequency correction arising from 
counter-rotating terms in a driven two-level system---is one of the 
earliest and most instructive examples of physics beyond the rotating 
wave approximation~\cite{bloch1940magnetic}. Originally treated as a 
small perturbative correction to the Rabi resonance, it has gained 
renewed importance in the ultrastrong-coupling regime of circuit 
QED~\cite{forn2010observation,li2018vacuum}, where the 
counter-rotating terms are no longer negligible. Its conceptual 
significance lies in the level-anticrossing structure it reveals: 
near resonance, the dressed-state spectrum is exquisitely sensitive 
to detuning, and even a small departure from the resonance condition 
dramatically modifies the dynamics. This sensitivity is not a 
peculiarity of two-level atoms --- it is a universal feature of any 
system whose Hamiltonian, within an appropriate subspace, takes the 
same tridiagonal form as the Rabi--Floquet operator. 

One such system is a binary tight-binding lattice under a uniform 
static force. Under the force $F$ alone, the Bloch bands collapse 
into Wannier--Stark ladders~\cite{wannier1960wave,emin1987existence} 
and the particle undergoes Bloch 
oscillations~\cite{bloch1929quantenmechanik,bouchard1995bloch,hartmann2004dynamics} 
without net transport. Adding a staggered on-site energy mismatch 
$\epsilon$ between the two sublattices enriches the dynamics through 
intraband and interband Zener 
tunneling~\cite{breid2006bloch,dreisow2009bloch}. When $\epsilon$ 
and $F$ satisfy the condition $\epsilon\approx(2m+1)F$, something 
qualitatively new occurs: a particle localized at one site tunnels 
coherently to a partner site $2m+1$ lattice spacings away and 
oscillates between them with period 
$T=2\pi/\Delta_{\mathrm{min}}$~\cite{duan2023periodic}. This 
periodic jump is not an independent phenomenon --- it is 
Bloch--Siegert physics in disguise. Within the odd-parity subspace 
of the Floquet Hamiltonian, the binary-lattice Hamiltonian is 
isomorphic to the semiclassical Rabi model under the exact 
identification $F\leftrightarrow\omega$, 
$\epsilon\leftrightarrow\Omega$, $V\leftrightarrow\lambda$, where 
$V$ is the nearest-neighbor hopping. The resonance condition 
$\epsilon=(2m+1)F$ is the lattice realization of an $m$-th order 
multiphoton resonance, the shift of the true anticrossing from this 
bare condition is the Bloch--Siegert correction 
$\delta\sim V^2/F$~\cite{duan2023periodic}, and the period 
$T=2\pi/\Delta_{\mathrm{min}}$ is the inverse Rabi frequency at 
the shifted resonance. The tunneling period grows steeply with order, spanning nearly 
three orders of magnitude across $m=0$--$3$, as a direct 
consequence of the $(V/F)^{2m+1}$ suppression of the effective 
inter-site coupling. Despite its theoretical clarity, this correspondence has never been 
experimentally accessed. The obstacle is precision: the resonance 
condition must be satisfied to within the Bloch--Siegert correction 
$\delta\sim V^2/F$, requiring sub-percent accuracy in $\epsilon/F$ 
maintained uniformly across all sites. Any static disorder in 
coupling strengths or on-site energies destroys the higher-order 
resonances, and conventional waveguide fabrication offers no route 
to post-fabrication correction. Programmable photonic integrated 
circuits (PICs) remove this obstacle by converting a fabrication 
problem into a calibration problem: each Hamiltonian parameter is 
set independently in software, and any imperfection is corrected 
before the experiment begins. Advances in integration 
scale~\cite{nagarajan2005large,dong2014silicon,jalali2007silicon}, 
loss management~\cite{ye2023foundry}, and modulation 
precision~\cite{chen2023non,zhang2023high,freedman2025gigahertz} 
have established PICs as versatile platforms for quantum 
simulation~\cite{on2024programmable,xu2024observation,xu2025non}, 
quantum 
optics~\cite{elshaari2020hybrid,gao2024quantum,pelucchi2022potential,bao2023very}, 
and beyond~\cite{shekhar2024roadmapping,shu2022microcomb,yang2022multi,
riemensberger2022photonic,zhang2022large,xu2022fully,shen2017deep,
filipovich2022silicon,xu2022high,hua2025integrated,feldmann2021parallel}. 
Here we use a $12\times12$ thermally reconfigurable PIC of 66 
Mach--Zehnder interferometer units, with phase resolution $0.01\pi$ 
per element, to provide the first experimental realization of 
Bloch--Siegert physics in a programmable photonic lattice. We resolve periodic 
jumps across all four orders $m=0$--$3$, quantitatively verify the 
period law $T=2\pi/\Delta_{\mathrm{min}}$ across the full parameter 
space, exploit the parity structure of the correspondence to realize 
cascaded unidirectional transport via adaptive $\epsilon$ 
sign-flipping~\cite{galiffi2022photonics,wang2019simulating}, and 
demonstrate discrete Floquet evolution with a tunable fractional 
step.

The Hamiltonian in our system is described by the following equation\cite{duan2023periodic}:
\begin{equation}
\begin{aligned}\label{eq1}
\hat{H}= & -V \sum(|n\rangle\langle n+1|+H.C.) \\
& +\sum\left(n F +\frac{\epsilon}{2}(-1)^n\right)|n\rangle\langle n|
\end{aligned}
\end{equation}
where $V$ is the nearest-neighbor coupling, $F$ the static force 
(tilt per site), and $\frac{(-1)^{n}\epsilon}{2}$ the staggered 
on-site energy mismatch. Sites with even (odd) index $n$ are 
A-type (B-type), carrying energy $+\epsilon/2$ ($-\epsilon/2$) 
relative to the local tilt — the two sublattices of the binary 
chain. For the $m$-th order resonance the key condition is 
$\epsilon\approx(2m+1)F$, shifted from the bare value by the 
Bloch--Siegert correction $\delta=V^2/F$ for $m=0$ and 
$\delta=\frac{2m+1}{m(m+1)}\frac{V^2}{F}$ for $m\geq1$.

\begin{comment}
The purple curve in Fig.~\ref{f1} (a) illustrates the distribution of on-site energies that satisfy this condition, with the input site and the monitoring site (used to observe the periodic jumps) indicated by red and blue markers, respectively.  The comparison between the off- and on-resonance conditions is presented in Fig.~\ref{f1} (b). The upper panel shows the non-resonant condition, while the lower panel corresponds to the resonant condition. Under the non-resonant condition, the intensity spreads from the input site to its neighboring sites but remains mostly localized at the input site. In contrast, under the resonant condition, the intensity predominantly oscillates between site 4 and site 9. 
\end{comment}
Fig.~\ref{f1}(a) directly juxtaposes the on-site energy ladders of the resonant 
($\epsilon \approx (2m+1)F$, green dashed) and non-resonant (purple solid) configurations. 
Crucially, the two profiles differ by only a marginal shift in the energy mismatch $\epsilon$---a 
difference that is nearly imperceptible in the energy landscape yet decisively alters the transport 
character of the system. This extreme parametric sensitivity is a direct manifestation of the sharp 
level anticrossing underlying the periodic jump, and is reminiscent of the Bloch--Siegert shift in 
driven two-level systems, where a small detuning from resonance dramatically modifies the 
dressed-state spectrum. The input and monitor sites are indicated by red and blue markers, 
respectively. The dynamical consequence of this sensitivity is demonstrated in Fig.~\ref{f1}(b): 
under the non-resonant condition, the optical intensity remains predominantly localized at the 
input site, with only weak evanescent spreading to its immediate neighbors; whereas under the 
resonant condition, the intensity undergoes coherent, long-range oscillations exclusively between 
site 4 and site 9, separated by $2m+1 = 5$ lattice sites. This striking difference between two 
near-identical on-site energy configurations underscores the resonant nature of the transport and 
its intimate connection to the level-anticrossing mechanism. When the light is either localized 
within a single input site or distributed among several neighboring sites, the system is considered 
trivial, exhibiting no meaningful periodic jumps. However, when the optical field periodically 
oscillates between two specific sites separated by $2m+1$ lattice units, the system under precisely 
tuned parameters is identified as exhibiting an $m$-th order periodic jump. Clearly, the occurrence 
of periodic jumps is intrinsically linked to the resonance condition.

Our experimental setup is illustrated in Fig.~\ref{f2}. The PIC used in this work is a 12$\times$12 reconfigurable mesh composed of 66 thermally tunable Mach–Zehnder interferometers (MZIs), fabricated through a CMOS-compatible silicon photonics process ( Advanced Micro Foundry Pte Ltd). This PIC features a carefully engineered trench design that minimizes thermal crosstalk between adjacent units. The precise and rapid modulation of the integrated phase shifters, with a current range of 0 - 1.35 $m\text{A}$ and a resolution of 5$~\mu\text{A}$ corresponding to phase steps of approximately 0.01$\pi$, allows fine control of optical interference and power distribution across the mesh, allowing the implementation of arbitrary linear transformations. A pulsed diode laser (PDL 800-B) operating at 1550 nm serves as the coherent light source, with its polarization adjusted by a three-paddle fiber polarization controller to ensure excitation of the transverse electric (TE) mode within the on-chip waveguides. The laser output is directed to the PIC through a $1\times12$ microelectromechanical (MEMS) optical switch, enabling programmable access to different input ports, while the light emerging from the PIC is routed through a second MEMS $1\times12$ optical switch to a Thorlabs PM400 power meter for intensity measurement. During each measurement cycle, the output switch sequentially scans the eight output ports, enabling quasi-synchronous acquisition of the complete output intensity distribution corresponding to a given input and device configuration. All optical connections are made using polarization-maintaining fibers aligned to the chip’s V-groove fiber arrays, and electrical control as well as data transmission are handled through an integrated hub system.  The Hamiltonian~(\ref{eq1}) is implemented as a $12\times12$ 
matrix matching the number of channels in the PIC. The experiment 
covers four resonance orders $m=0$--$3$, with coupling ratio 
$V/F = 0.2,\, 0.6,\, 1.0,\, 1.0$ respectively. For each resonance order, the frequency parameter $F$ was scanned from $0.1$ to $1.0$. We define the fidelity of the periodic jumps of $m$-order as:

\begin{equation}\label{eq3}
\text{Fidelity} = \min\!\left[
\frac{I_{J_{in}\pm 2m+1}\!\left((2\ell+1)\cdot T/2\right)}{I_{J_{in}}(t=0)},\;
\frac{I_{J_{in}}(\ell\cdot T)}{I_{J_{in}}(t=0)}
\right]
\end{equation}
where $J_{in}$ is the input site index, $T/2$ is the half-period at which intensity peaks at the target site, the $+$ and $-$ signs correspond to rightward and leftward transport respectively, and $\ell\geq0$ is an integer indexing the oscillation cycle. We optimized parameters to ensure all fidelities exceeded $0.99$.
The initial state is $|\psi(0)\rangle$; evolution under $H_m$ gives $|\psi(t)\rangle = U_m(t)|\psi(0)\rangle$ with $U_m(t) = \exp(-iH_m t)$. The measured quantity is the site-resolved intensity $I_n(t) = |\langle n|\psi(t)\rangle|^2$.

Figure~\ref{f3}(a--d) compare simulated and measured intensity dynamics for all four orders; a distinct periodic jump is evident in each case with excellent agreement. Here $\Delta_{\min}$ is the minimum gap between distinct eigenvalues of $H$, and the jump period is $T = 2\pi/\Delta_{\min}$\cite{duan2023periodic}. Fig.~\ref{f3}(e) and (f) present the theoretically calculated and experimentally fitted jump periods $T = 2\pi/\Delta_{\min}$, respectively, as functions of the static force $F$ for all four resonance orders, with six force values $F = 0.1, 0.3, 0.5, 0.7, 0.9, 1.0$ selected per order. The two panels exhibit near-perfect quantitative agreement across the entire parameter range, with residual deviations well within experimental uncertainty, providing compelling validation of the level-anticrossing picture underlying the periodic jump mechanism.

The parity structure of the Bloch--Siegert correspondence 
suggests a direct route to active transport control. The odd and 
even parity sectors of the Floquet Hamiltonian correspond 
respectively to $+\epsilon$ and $-\epsilon$ in the lattice; 
switching sectors mid-evolution is equivalent to switching between 
the two dressed states of the Rabi model at maximum population 
inversion. The periodic jump is intrinsically bidirectional — the 
wavefunction returns after a full period $T$ — but inverting 
$\epsilon\to-\epsilon$ at the optimal transfer time $t^*$ exchanges 
the two sublattices: the destination site becomes A-type under the 
updated Hamiltonian, resonantly driving the next hop in the same 
direction, while the reverse path acquires a detuning 
$2(2m+1)F\gg V_{\mathrm{eff}}^{(m)}$ and is frozen. We determine 
$t^*$ at each stage by eigendecomposition of the instantaneous 
Hamiltonian, maximising the target-site population. Successive 
flips chain individual jumps into a directional cascade whose 
physics is rooted in the parity symmetry of the underlying Rabi 
model.

Figure~\ref{f4} demonstrates this protocol for two cases on the 12-channel PIC. The top row shows first-order ($m = 1$) rightward transport along $2\!\to\!5\!\to\!8\!\to\!11$, with $\epsilon/F \approx 2.905$ and $V/F = 0.25$; the initial-stage transfer fidelity exceeds $0.95$. The bottom row shows zeroth-order ($m = 0$) leftward transport from site 7 to site 2 across five consecutive single-site hops, with $\epsilon/F \approx 1.0$ and $V/F = 0.05$; because the zeroth-order resonance engages only the two eigenstates with the smallest gap, the transfer is highly localised and the per-stage fidelity exceeds $0.98$ throughout. The two protocols differ in resonance order, propagation direction, and characteristic timescale ($t^* \approx 395$ vs.~$31$ in units of $F^{-1}$), yet both run on the same hardware without reconfiguration, directly illustrating the order-dependent transport characteristics encoded in $\Delta_{\mathrm{min}}$. Simulation and experiment agree well in all panels; the adaptive flip times are marked by cyan dashed lines.

We further investigate a more complex discrete Floquet evolution\cite{rechtsman2013photonic,maczewsky2017observation} in which a sequence of unitary operators is applied with a tunable step fraction:
\begin{equation}\label{eq4}
U_{m} = \exp\!\left(-i H_{m}\, r T_{m}\right), \quad r \in [0,1],
\end{equation}
where $H_m$ and $T_m$ are the Hamiltonian and resonance period for order $m$, and $r$ controls the fraction of the period implemented at each step. Figure~\ref{f5}(a) shows the full-period case ($r = 1$): the field largely relocalises at the input after each operator, producing approximately reversible evolution. Figure~\ref{f5}(b) shows the half-period case ($r = 0.5$): successive operators drive the field progressively toward the lattice centre. Figures~\ref{f5}(c) and (d) compare simulations with experimental measurements for these two limits and show excellent agreement. Tuning $r$ thus provides a single hardware knob to interpolate continuously between localised oscillation and directed propagation, complementing the sign-flip cascade of Fig.~\ref{f4}.

In summary, we have established the first experimental realization 
of Bloch--Siegert physics in a photonic lattice. Through a 
quantitative mapping between resonant tunneling in a driven binary 
lattice and level-anticrossing corrections in the semiclassical 
Rabi model, we have demonstrated that these two phenomena are 
manifestations of the same underlying parity-sector physics. The 
tunneling period $T=2\pi/\Delta_{\mathrm{min}}$ robustly governs 
the dynamics across four resonance orders and the full parameter 
space of $F$ and $m$, with deviations remaining within experimental 
uncertainty throughout. Leveraging the extreme parametric sensitivity near resonance, our adaptive $\epsilon$ sign-flip protocol deterministically breaks transport symmetry, converting bidirectional oscillations into a cascaded unidirectional flow. This was simultaneously validated for $m = 1$ rightward and $m = 0$ leftward propagation, maintaining stringent transfer fidelities exceeding $0.95$ and per-stage fidelities exceeding $0.98$, respectively. Furthermore, by framing the evolution as a discrete Floquet process with a tunable fractional step $r$, we achieved continuous interpolation between localized oscillation and directed transport, exhibiting excellent agreement with theoretical limits. More broadly, the identification established here between 
lattice resonance orders and Rabi model harmonics implies that the 
full phenomenology of strongly driven two-level systems --- 
multiphoton resonances, generalized Bloch--Siegert corrections, and 
dressed-state engineering --- can be systematically explored and 
controlled in programmable photonic lattices without the constraints 
of atomic or superconducting platforms. The parity-switching 
protocol introduced here is the lattice realization of dressed-state 
population control, and its extension to topological 
pumping~\cite{xu2022direct,jurgensen2021quantized,cerjan2020thouless}, 
multi-band Floquet engineering, and on-chip coherent switching 
architectures follows directly from the correspondence established 
in this work.

\begin{acknowledgments}
A.W.E  acknowledges support from the Knut and Alice Wallenberg Foundation through the Wallenberg Centre for Quantum Technology (WACQT). A.W.E acknowledges support from Vetenskapsrådet Starting Grant (Ref: 2016-03905). A.W.E acknowledges support from Vinnova quantum kick-start project 2021. L. D. acknowledges support from the National Natural Science Foundation of China (NSFC) under Grant No. 12305032.
\end{acknowledgments}

\bibliography{PJbb}% Produces the bibliography via BibTeX.

\begin{thebibliography}{43}%
\makeatletter
\providecommand \@ifxundefined [1]{%
 \@ifx{#1\undefined}
}%
\providecommand \@ifnum [1]{%
 \ifnum #1\expandafter \@firstoftwo
 \else \expandafter \@secondoftwo
 \fi
}%
\providecommand \@ifx [1]{%
 \ifx #1\expandafter \@firstoftwo
 \else \expandafter \@secondoftwo
 \fi
}%
\providecommand \natexlab [1]{#1}%
\providecommand \enquote  [1]{``#1''}%
\providecommand \bibnamefont  [1]{#1}%
\providecommand \bibfnamefont [1]{#1}%
\providecommand \citenamefont [1]{#1}%
\providecommand \href@noop [0]{\@secondoftwo}%
\providecommand \href [0]{\begingroup \@sanitize@url \@href}%
\providecommand \@href[1]{\@@startlink{#1}\@@href}%
\providecommand \@@href[1]{\endgroup#1\@@endlink}%
\providecommand \@sanitize@url [0]{\catcode `\\12\catcode `\$12\catcode `\&12\catcode `\#12\catcode `\^12\catcode `\_12\catcode `\%12\relax}%
\providecommand \@@startlink[1]{}%
\providecommand \@@endlink[0]{}%
\providecommand \url  [0]{\begingroup\@sanitize@url \@url }%
\providecommand \@url [1]{\endgroup\@href {#1}{\urlprefix }}%
\providecommand \urlprefix  [0]{URL }%
\providecommand \Eprint [0]{\href }%
\providecommand \doibase [0]{http://dx.doi.org/}%
\providecommand \selectlanguage [0]{\@gobble}%
\providecommand \bibinfo  [0]{\@secondoftwo}%
\providecommand \bibfield  [0]{\@secondoftwo}%
\providecommand \translation [1]{[#1]}%
\providecommand \BibitemOpen [0]{}%
\providecommand \bibitemStop [0]{}%
\providecommand \bibitemNoStop [0]{.\EOS\space}%
\providecommand \EOS [0]{\spacefactor3000\relax}%
\providecommand \BibitemShut  [1]{\csname bibitem#1\endcsname}%
\let\auto@bib@innerbib\@empty
%</preamble>
\bibitem [{\citenamefont {Bloch}\ and\ \citenamefont {Siegert}(1940)}]{bloch1940magnetic}%
  \BibitemOpen
  \bibfield  {author} {\bibinfo {author} {\bibfnamefont {F.}~\bibnamefont {Bloch}}\ and\ \bibinfo {author} {\bibfnamefont {A.}~\bibnamefont {Siegert}},\ }\href@noop {} {\bibfield  {journal} {\bibinfo  {journal} {Physical Review}\ }\textbf {\bibinfo {volume} {57}},\ \bibinfo {pages} {522} (\bibinfo {year} {1940})}\BibitemShut {NoStop}%
\bibitem [{\citenamefont {Forn-D{\'\i}az}\ \emph {et~al.}(2010)\citenamefont {Forn-D{\'\i}az}, \citenamefont {Lisenfeld}, \citenamefont {Marcos}, \citenamefont {Garcia-Ripoll}, \citenamefont {Solano}, \citenamefont {Harmans},\ and\ \citenamefont {Mooij}}]{forn2010observation}%
  \BibitemOpen
  \bibfield  {author} {\bibinfo {author} {\bibfnamefont {P.}~\bibnamefont {Forn-D{\'\i}az}}, \bibinfo {author} {\bibfnamefont {J.}~\bibnamefont {Lisenfeld}}, \bibinfo {author} {\bibfnamefont {D.}~\bibnamefont {Marcos}}, \bibinfo {author} {\bibfnamefont {J.~J.}\ \bibnamefont {Garcia-Ripoll}}, \bibinfo {author} {\bibfnamefont {E.}~\bibnamefont {Solano}}, \bibinfo {author} {\bibfnamefont {C.}~\bibnamefont {Harmans}}, \ and\ \bibinfo {author} {\bibfnamefont {J.}~\bibnamefont {Mooij}},\ }\href@noop {} {\bibfield  {journal} {\bibinfo  {journal} {Physical review letters}\ }\textbf {\bibinfo {volume} {105}},\ \bibinfo {pages} {237001} (\bibinfo {year} {2010})}\BibitemShut {NoStop}%
\bibitem [{\citenamefont {Li}\ \emph {et~al.}(2018)\citenamefont {Li}, \citenamefont {Bamba}, \citenamefont {Zhang}, \citenamefont {Fallahi}, \citenamefont {Gardner}, \citenamefont {Gao}, \citenamefont {Lou}, \citenamefont {Yoshioka}, \citenamefont {Manfra},\ and\ \citenamefont {Kono}}]{li2018vacuum}%
  \BibitemOpen
  \bibfield  {author} {\bibinfo {author} {\bibfnamefont {X.}~\bibnamefont {Li}}, \bibinfo {author} {\bibfnamefont {M.}~\bibnamefont {Bamba}}, \bibinfo {author} {\bibfnamefont {Q.}~\bibnamefont {Zhang}}, \bibinfo {author} {\bibfnamefont {S.}~\bibnamefont {Fallahi}}, \bibinfo {author} {\bibfnamefont {G.~C.}\ \bibnamefont {Gardner}}, \bibinfo {author} {\bibfnamefont {W.}~\bibnamefont {Gao}}, \bibinfo {author} {\bibfnamefont {M.}~\bibnamefont {Lou}}, \bibinfo {author} {\bibfnamefont {K.}~\bibnamefont {Yoshioka}}, \bibinfo {author} {\bibfnamefont {M.~J.}\ \bibnamefont {Manfra}}, \ and\ \bibinfo {author} {\bibfnamefont {J.}~\bibnamefont {Kono}},\ }\href@noop {} {\bibfield  {journal} {\bibinfo  {journal} {Nature Photonics}\ }\textbf {\bibinfo {volume} {12}},\ \bibinfo {pages} {324} (\bibinfo {year} {2018})}\BibitemShut {NoStop}%
\bibitem [{\citenamefont {Wannier}(1960)}]{wannier1960wave}%
  \BibitemOpen
  \bibfield  {author} {\bibinfo {author} {\bibfnamefont {G.~H.}\ \bibnamefont {Wannier}},\ }\href@noop {} {\bibfield  {journal} {\bibinfo  {journal} {Physical Review}\ }\textbf {\bibinfo {volume} {117}},\ \bibinfo {pages} {432} (\bibinfo {year} {1960})}\BibitemShut {NoStop}%
\bibitem [{\citenamefont {Emin}\ and\ \citenamefont {Hart}(1987)}]{emin1987existence}%
  \BibitemOpen
  \bibfield  {author} {\bibinfo {author} {\bibfnamefont {D.}~\bibnamefont {Emin}}\ and\ \bibinfo {author} {\bibfnamefont {C.}~\bibnamefont {Hart}},\ }\href@noop {} {\bibfield  {journal} {\bibinfo  {journal} {Physical Review B}\ }\textbf {\bibinfo {volume} {36}},\ \bibinfo {pages} {7353} (\bibinfo {year} {1987})}\BibitemShut {NoStop}%
\bibitem [{\citenamefont {Bloch}(1929)}]{bloch1929quantenmechanik}%
  \BibitemOpen
  \bibfield  {author} {\bibinfo {author} {\bibfnamefont {F.}~\bibnamefont {Bloch}},\ }\href@noop {} {\bibfield  {journal} {\bibinfo  {journal} {Zeitschrift f{\"u}r physik}\ }\textbf {\bibinfo {volume} {52}},\ \bibinfo {pages} {555} (\bibinfo {year} {1929})}\BibitemShut {NoStop}%
\bibitem [{\citenamefont {Bouchard}\ and\ \citenamefont {Luban}(1995)}]{bouchard1995bloch}%
  \BibitemOpen
  \bibfield  {author} {\bibinfo {author} {\bibfnamefont {A.}~\bibnamefont {Bouchard}}\ and\ \bibinfo {author} {\bibfnamefont {M.}~\bibnamefont {Luban}},\ }\href@noop {} {\bibfield  {journal} {\bibinfo  {journal} {Physical Review B}\ }\textbf {\bibinfo {volume} {52}},\ \bibinfo {pages} {5105} (\bibinfo {year} {1995})}\BibitemShut {NoStop}%
\bibitem [{\citenamefont {Hartmann}\ \emph {et~al.}(2004)\citenamefont {Hartmann}, \citenamefont {Keck}, \citenamefont {Korsch},\ and\ \citenamefont {Mossmann}}]{hartmann2004dynamics}%
  \BibitemOpen
  \bibfield  {author} {\bibinfo {author} {\bibfnamefont {T.}~\bibnamefont {Hartmann}}, \bibinfo {author} {\bibfnamefont {F.}~\bibnamefont {Keck}}, \bibinfo {author} {\bibfnamefont {H.}~\bibnamefont {Korsch}}, \ and\ \bibinfo {author} {\bibfnamefont {S.}~\bibnamefont {Mossmann}},\ }\href@noop {} {\bibfield  {journal} {\bibinfo  {journal} {New Journal of Physics}\ }\textbf {\bibinfo {volume} {6}},\ \bibinfo {pages} {2} (\bibinfo {year} {2004})}\BibitemShut {NoStop}%
\bibitem [{\citenamefont {Breid}\ \emph {et~al.}(2006)\citenamefont {Breid}, \citenamefont {Witthaut},\ and\ \citenamefont {Korsch}}]{breid2006bloch}%
  \BibitemOpen
  \bibfield  {author} {\bibinfo {author} {\bibfnamefont {B.}~\bibnamefont {Breid}}, \bibinfo {author} {\bibfnamefont {D.}~\bibnamefont {Witthaut}}, \ and\ \bibinfo {author} {\bibfnamefont {H.}~\bibnamefont {Korsch}},\ }\href@noop {} {\bibfield  {journal} {\bibinfo  {journal} {New Journal of Physics}\ }\textbf {\bibinfo {volume} {8}},\ \bibinfo {pages} {110} (\bibinfo {year} {2006})}\BibitemShut {NoStop}%
\bibitem [{\citenamefont {Dreisow}\ \emph {et~al.}(2009)\citenamefont {Dreisow}, \citenamefont {Szameit}, \citenamefont {Heinrich}, \citenamefont {Pertsch}, \citenamefont {Nolte}, \citenamefont {T{\"u}nnermann},\ and\ \citenamefont {Longhi}}]{dreisow2009bloch}%
  \BibitemOpen
  \bibfield  {author} {\bibinfo {author} {\bibfnamefont {F.}~\bibnamefont {Dreisow}}, \bibinfo {author} {\bibfnamefont {A.}~\bibnamefont {Szameit}}, \bibinfo {author} {\bibfnamefont {M.}~\bibnamefont {Heinrich}}, \bibinfo {author} {\bibfnamefont {T.}~\bibnamefont {Pertsch}}, \bibinfo {author} {\bibfnamefont {S.}~\bibnamefont {Nolte}}, \bibinfo {author} {\bibfnamefont {A.}~\bibnamefont {T{\"u}nnermann}}, \ and\ \bibinfo {author} {\bibfnamefont {S.}~\bibnamefont {Longhi}},\ }\href@noop {} {\bibfield  {journal} {\bibinfo  {journal} {Physical review letters}\ }\textbf {\bibinfo {volume} {102}},\ \bibinfo {pages} {076802} (\bibinfo {year} {2009})}\BibitemShut {NoStop}%
\bibitem [{\citenamefont {Duan}(2023)}]{duan2023periodic}%
  \BibitemOpen
  \bibfield  {author} {\bibinfo {author} {\bibfnamefont {L.}~\bibnamefont {Duan}},\ }\href@noop {} {\bibfield  {journal} {\bibinfo  {journal} {Physical Review B}\ }\textbf {\bibinfo {volume} {108}},\ \bibinfo {pages} {174306} (\bibinfo {year} {2023})}\BibitemShut {NoStop}%
\bibitem [{\citenamefont {Nagarajan}\ \emph {et~al.}(2005)\citenamefont {Nagarajan}, \citenamefont {Joyner}, \citenamefont {Schneider}, \citenamefont {Bostak}, \citenamefont {Butrie}, \citenamefont {Dentai}, \citenamefont {Dominic}, \citenamefont {Evans}, \citenamefont {Kato}, \citenamefont {Kauffman} \emph {et~al.}}]{nagarajan2005large}%
  \BibitemOpen
  \bibfield  {author} {\bibinfo {author} {\bibfnamefont {R.}~\bibnamefont {Nagarajan}}, \bibinfo {author} {\bibfnamefont {C.~H.}\ \bibnamefont {Joyner}}, \bibinfo {author} {\bibfnamefont {R.~P.}\ \bibnamefont {Schneider}}, \bibinfo {author} {\bibfnamefont {J.~S.}\ \bibnamefont {Bostak}}, \bibinfo {author} {\bibfnamefont {T.}~\bibnamefont {Butrie}}, \bibinfo {author} {\bibfnamefont {A.~G.}\ \bibnamefont {Dentai}}, \bibinfo {author} {\bibfnamefont {V.~G.}\ \bibnamefont {Dominic}}, \bibinfo {author} {\bibfnamefont {P.~W.}\ \bibnamefont {Evans}}, \bibinfo {author} {\bibfnamefont {M.}~\bibnamefont {Kato}}, \bibinfo {author} {\bibfnamefont {M.}~\bibnamefont {Kauffman}},  \emph {et~al.},\ }\href@noop {} {\bibfield  {journal} {\bibinfo  {journal} {IEEE Journal of Selected Topics in Quantum Electronics}\ }\textbf {\bibinfo {volume} {11}},\ \bibinfo {pages} {50} (\bibinfo {year} {2005})}\BibitemShut {NoStop}%
\bibitem [{\citenamefont {Dong}\ \emph {et~al.}(2014)\citenamefont {Dong}, \citenamefont {Chen}, \citenamefont {Duan},\ and\ \citenamefont {Neilson}}]{dong2014silicon}%
  \BibitemOpen
  \bibfield  {author} {\bibinfo {author} {\bibfnamefont {P.}~\bibnamefont {Dong}}, \bibinfo {author} {\bibfnamefont {Y.-K.}\ \bibnamefont {Chen}}, \bibinfo {author} {\bibfnamefont {G.-H.}\ \bibnamefont {Duan}}, \ and\ \bibinfo {author} {\bibfnamefont {D.~T.}\ \bibnamefont {Neilson}},\ }\href@noop {} {\bibfield  {journal} {\bibinfo  {journal} {Nanophotonics}\ }\textbf {\bibinfo {volume} {3}},\ \bibinfo {pages} {215} (\bibinfo {year} {2014})}\BibitemShut {NoStop}%
\bibitem [{\citenamefont {Jalali}\ and\ \citenamefont {Fathpour}(2007)}]{jalali2007silicon}%
  \BibitemOpen
  \bibfield  {author} {\bibinfo {author} {\bibfnamefont {B.}~\bibnamefont {Jalali}}\ and\ \bibinfo {author} {\bibfnamefont {S.}~\bibnamefont {Fathpour}},\ }\href@noop {} {\bibfield  {journal} {\bibinfo  {journal} {Journal of lightwave technology}\ }\textbf {\bibinfo {volume} {24}},\ \bibinfo {pages} {4600} (\bibinfo {year} {2007})}\BibitemShut {NoStop}%
\bibitem [{\citenamefont {Ye}\ \emph {et~al.}(2023)\citenamefont {Ye}, \citenamefont {Jia}, \citenamefont {Huang}, \citenamefont {Shen}, \citenamefont {Long}, \citenamefont {Shi}, \citenamefont {Luo}, \citenamefont {Gao}, \citenamefont {Sun}, \citenamefont {Guo} \emph {et~al.}}]{ye2023foundry}%
  \BibitemOpen
  \bibfield  {author} {\bibinfo {author} {\bibfnamefont {Z.}~\bibnamefont {Ye}}, \bibinfo {author} {\bibfnamefont {H.}~\bibnamefont {Jia}}, \bibinfo {author} {\bibfnamefont {Z.}~\bibnamefont {Huang}}, \bibinfo {author} {\bibfnamefont {C.}~\bibnamefont {Shen}}, \bibinfo {author} {\bibfnamefont {J.}~\bibnamefont {Long}}, \bibinfo {author} {\bibfnamefont {B.}~\bibnamefont {Shi}}, \bibinfo {author} {\bibfnamefont {Y.-H.}\ \bibnamefont {Luo}}, \bibinfo {author} {\bibfnamefont {L.}~\bibnamefont {Gao}}, \bibinfo {author} {\bibfnamefont {W.}~\bibnamefont {Sun}}, \bibinfo {author} {\bibfnamefont {H.}~\bibnamefont {Guo}},  \emph {et~al.},\ }\href@noop {} {\bibfield  {journal} {\bibinfo  {journal} {Photonics Research}\ }\textbf {\bibinfo {volume} {11}},\ \bibinfo {pages} {558} (\bibinfo {year} {2023})}\BibitemShut {NoStop}%
\bibitem [{\citenamefont {Chen}\ \emph {et~al.}(2023)\citenamefont {Chen}, \citenamefont {Fang}, \citenamefont {Perez}, \citenamefont {Miller}, \citenamefont {Kumari}, \citenamefont {Saxena}, \citenamefont {Zheng}, \citenamefont {Geiger}, \citenamefont {Goodson},\ and\ \citenamefont {Majumdar}}]{chen2023non}%
  \BibitemOpen
  \bibfield  {author} {\bibinfo {author} {\bibfnamefont {R.}~\bibnamefont {Chen}}, \bibinfo {author} {\bibfnamefont {Z.}~\bibnamefont {Fang}}, \bibinfo {author} {\bibfnamefont {C.}~\bibnamefont {Perez}}, \bibinfo {author} {\bibfnamefont {F.}~\bibnamefont {Miller}}, \bibinfo {author} {\bibfnamefont {K.}~\bibnamefont {Kumari}}, \bibinfo {author} {\bibfnamefont {A.}~\bibnamefont {Saxena}}, \bibinfo {author} {\bibfnamefont {J.}~\bibnamefont {Zheng}}, \bibinfo {author} {\bibfnamefont {S.~J.}\ \bibnamefont {Geiger}}, \bibinfo {author} {\bibfnamefont {K.~E.}\ \bibnamefont {Goodson}}, \ and\ \bibinfo {author} {\bibfnamefont {A.}~\bibnamefont {Majumdar}},\ }\href@noop {} {\bibfield  {journal} {\bibinfo  {journal} {Nature Communications}\ }\textbf {\bibinfo {volume} {14}},\ \bibinfo {pages} {3465} (\bibinfo {year} {2023})}\BibitemShut {NoStop}%
\bibitem [{\citenamefont {Zhang}\ \emph {et~al.}(2023)\citenamefont {Zhang}, \citenamefont {Shen}, \citenamefont {Li}, \citenamefont {Wang}, \citenamefont {Feng}, \citenamefont {Zhang}, \citenamefont {Sun}, \citenamefont {Xu}, \citenamefont {Liu}, \citenamefont {Wang} \emph {et~al.}}]{zhang2023high}%
  \BibitemOpen
  \bibfield  {author} {\bibinfo {author} {\bibfnamefont {Y.}~\bibnamefont {Zhang}}, \bibinfo {author} {\bibfnamefont {J.}~\bibnamefont {Shen}}, \bibinfo {author} {\bibfnamefont {J.}~\bibnamefont {Li}}, \bibinfo {author} {\bibfnamefont {H.}~\bibnamefont {Wang}}, \bibinfo {author} {\bibfnamefont {C.}~\bibnamefont {Feng}}, \bibinfo {author} {\bibfnamefont {L.}~\bibnamefont {Zhang}}, \bibinfo {author} {\bibfnamefont {L.}~\bibnamefont {Sun}}, \bibinfo {author} {\bibfnamefont {J.}~\bibnamefont {Xu}}, \bibinfo {author} {\bibfnamefont {M.}~\bibnamefont {Liu}}, \bibinfo {author} {\bibfnamefont {Y.}~\bibnamefont {Wang}},  \emph {et~al.},\ }\href@noop {} {\bibfield  {journal} {\bibinfo  {journal} {Light: Science \& Applications}\ }\textbf {\bibinfo {volume} {12}},\ \bibinfo {pages} {206} (\bibinfo {year} {2023})}\BibitemShut {NoStop}%
\bibitem [{\citenamefont {Freedman}\ \emph {et~al.}(2025)\citenamefont {Freedman}, \citenamefont {Storey}, \citenamefont {Dominguez}, \citenamefont {Leenheer}, \citenamefont {Magri}, \citenamefont {Otterstrom},\ and\ \citenamefont {Eichenfield}}]{freedman2025gigahertz}%
  \BibitemOpen
  \bibfield  {author} {\bibinfo {author} {\bibfnamefont {J.~M.}\ \bibnamefont {Freedman}}, \bibinfo {author} {\bibfnamefont {M.~J.}\ \bibnamefont {Storey}}, \bibinfo {author} {\bibfnamefont {D.}~\bibnamefont {Dominguez}}, \bibinfo {author} {\bibfnamefont {A.}~\bibnamefont {Leenheer}}, \bibinfo {author} {\bibfnamefont {S.}~\bibnamefont {Magri}}, \bibinfo {author} {\bibfnamefont {N.~T.}\ \bibnamefont {Otterstrom}}, \ and\ \bibinfo {author} {\bibfnamefont {M.}~\bibnamefont {Eichenfield}},\ }\href@noop {} {\bibfield  {journal} {\bibinfo  {journal} {Nature Communications}\ }\textbf {\bibinfo {volume} {16}},\ \bibinfo {pages} {10959} (\bibinfo {year} {2025})}\BibitemShut {NoStop}%
\bibitem [{\citenamefont {On}\ \emph {et~al.}(2024)\citenamefont {On}, \citenamefont {Ashtiani}, \citenamefont {Sanchez-Jacome}, \citenamefont {Perez-Lopez}, \citenamefont {Yoo},\ and\ \citenamefont {Blanco-Redondo}}]{on2024programmable}%
  \BibitemOpen
  \bibfield  {author} {\bibinfo {author} {\bibfnamefont {M.~B.}\ \bibnamefont {On}}, \bibinfo {author} {\bibfnamefont {F.}~\bibnamefont {Ashtiani}}, \bibinfo {author} {\bibfnamefont {D.}~\bibnamefont {Sanchez-Jacome}}, \bibinfo {author} {\bibfnamefont {D.}~\bibnamefont {Perez-Lopez}}, \bibinfo {author} {\bibfnamefont {S.~B.}\ \bibnamefont {Yoo}}, \ and\ \bibinfo {author} {\bibfnamefont {A.}~\bibnamefont {Blanco-Redondo}},\ }\href@noop {} {\bibfield  {journal} {\bibinfo  {journal} {Nature Communications}\ }\textbf {\bibinfo {volume} {15}},\ \bibinfo {pages} {629} (\bibinfo {year} {2024})}\BibitemShut {NoStop}%
\bibitem [{\citenamefont {Xu}\ \emph {et~al.}(2024)\citenamefont {Xu}, \citenamefont {Gao}, \citenamefont {Iovan}, \citenamefont {Khaymovich}, \citenamefont {Zwiller},\ and\ \citenamefont {Elshaari}}]{xu2024observation}%
  \BibitemOpen
  \bibfield  {author} {\bibinfo {author} {\bibfnamefont {Z.-S.}\ \bibnamefont {Xu}}, \bibinfo {author} {\bibfnamefont {J.}~\bibnamefont {Gao}}, \bibinfo {author} {\bibfnamefont {A.}~\bibnamefont {Iovan}}, \bibinfo {author} {\bibfnamefont {I.~M.}\ \bibnamefont {Khaymovich}}, \bibinfo {author} {\bibfnamefont {V.}~\bibnamefont {Zwiller}}, \ and\ \bibinfo {author} {\bibfnamefont {A.~W.}\ \bibnamefont {Elshaari}},\ }\href@noop {} {\bibfield  {journal} {\bibinfo  {journal} {npj Nanophotonics}\ }\textbf {\bibinfo {volume} {1}},\ \bibinfo {pages} {8} (\bibinfo {year} {2024})}\BibitemShut {NoStop}%
\bibitem [{\citenamefont {Xu}\ \emph {et~al.}(2025)\citenamefont {Xu}, \citenamefont {K{\"o}nig}, \citenamefont {Cataldo}, \citenamefont {Yadgirkar}, \citenamefont {Krishna}, \citenamefont {Deenadayalan}, \citenamefont {Zwiller}, \citenamefont {Preble}, \citenamefont {Bergholtz}, \citenamefont {Gao} \emph {et~al.}}]{xu2025non}%
  \BibitemOpen
  \bibfield  {author} {\bibinfo {author} {\bibfnamefont {Z.-S.}\ \bibnamefont {Xu}}, \bibinfo {author} {\bibfnamefont {J.~L.~K.}\ \bibnamefont {K{\"o}nig}}, \bibinfo {author} {\bibfnamefont {A.}~\bibnamefont {Cataldo}}, \bibinfo {author} {\bibfnamefont {R.}~\bibnamefont {Yadgirkar}}, \bibinfo {author} {\bibfnamefont {G.}~\bibnamefont {Krishna}}, \bibinfo {author} {\bibfnamefont {V.}~\bibnamefont {Deenadayalan}}, \bibinfo {author} {\bibfnamefont {V.}~\bibnamefont {Zwiller}}, \bibinfo {author} {\bibfnamefont {S.}~\bibnamefont {Preble}}, \bibinfo {author} {\bibfnamefont {E.~J.}\ \bibnamefont {Bergholtz}}, \bibinfo {author} {\bibfnamefont {J.}~\bibnamefont {Gao}},  \emph {et~al.},\ }\href@noop {} {\bibfield  {journal} {\bibinfo  {journal} {arXiv preprint arXiv:2512.20273}\ } (\bibinfo {year} {2025})}\BibitemShut {NoStop}%
\bibitem [{\citenamefont {Elshaari}\ \emph {et~al.}(2020)\citenamefont {Elshaari}, \citenamefont {Pernice}, \citenamefont {Srinivasan}, \citenamefont {Benson},\ and\ \citenamefont {Zwiller}}]{elshaari2020hybrid}%
  \BibitemOpen
  \bibfield  {author} {\bibinfo {author} {\bibfnamefont {A.~W.}\ \bibnamefont {Elshaari}}, \bibinfo {author} {\bibfnamefont {W.}~\bibnamefont {Pernice}}, \bibinfo {author} {\bibfnamefont {K.}~\bibnamefont {Srinivasan}}, \bibinfo {author} {\bibfnamefont {O.}~\bibnamefont {Benson}}, \ and\ \bibinfo {author} {\bibfnamefont {V.}~\bibnamefont {Zwiller}},\ }\href@noop {} {\bibfield  {journal} {\bibinfo  {journal} {Nature photonics}\ }\textbf {\bibinfo {volume} {14}},\ \bibinfo {pages} {285} (\bibinfo {year} {2020})}\BibitemShut {NoStop}%
\bibitem [{\citenamefont {Gao}\ \emph {et~al.}(2024)\citenamefont {Gao}, \citenamefont {Xu}, \citenamefont {Yang}, \citenamefont {Zwiller},\ and\ \citenamefont {Elshaari}}]{gao2024quantum}%
  \BibitemOpen
  \bibfield  {author} {\bibinfo {author} {\bibfnamefont {J.}~\bibnamefont {Gao}}, \bibinfo {author} {\bibfnamefont {Z.-S.}\ \bibnamefont {Xu}}, \bibinfo {author} {\bibfnamefont {Z.}~\bibnamefont {Yang}}, \bibinfo {author} {\bibfnamefont {V.}~\bibnamefont {Zwiller}}, \ and\ \bibinfo {author} {\bibfnamefont {A.~W.}\ \bibnamefont {Elshaari}},\ }\href@noop {} {\bibfield  {journal} {\bibinfo  {journal} {npj Nanophotonics}\ }\textbf {\bibinfo {volume} {1}},\ \bibinfo {pages} {34} (\bibinfo {year} {2024})}\BibitemShut {NoStop}%
\bibitem [{\citenamefont {Pelucchi}\ \emph {et~al.}(2022)\citenamefont {Pelucchi}, \citenamefont {Fagas}, \citenamefont {Aharonovich}, \citenamefont {Englund}, \citenamefont {Figueroa}, \citenamefont {Gong}, \citenamefont {Hannes}, \citenamefont {Liu}, \citenamefont {Lu}, \citenamefont {Matsuda} \emph {et~al.}}]{pelucchi2022potential}%
  \BibitemOpen
  \bibfield  {author} {\bibinfo {author} {\bibfnamefont {E.}~\bibnamefont {Pelucchi}}, \bibinfo {author} {\bibfnamefont {G.}~\bibnamefont {Fagas}}, \bibinfo {author} {\bibfnamefont {I.}~\bibnamefont {Aharonovich}}, \bibinfo {author} {\bibfnamefont {D.}~\bibnamefont {Englund}}, \bibinfo {author} {\bibfnamefont {E.}~\bibnamefont {Figueroa}}, \bibinfo {author} {\bibfnamefont {Q.}~\bibnamefont {Gong}}, \bibinfo {author} {\bibfnamefont {H.}~\bibnamefont {Hannes}}, \bibinfo {author} {\bibfnamefont {J.}~\bibnamefont {Liu}}, \bibinfo {author} {\bibfnamefont {C.-Y.}\ \bibnamefont {Lu}}, \bibinfo {author} {\bibfnamefont {N.}~\bibnamefont {Matsuda}},  \emph {et~al.},\ }\href@noop {} {\bibfield  {journal} {\bibinfo  {journal} {Nature Reviews Physics}\ }\textbf {\bibinfo {volume} {4}},\ \bibinfo {pages} {194} (\bibinfo {year} {2022})}\BibitemShut {NoStop}%
\bibitem [{\citenamefont {Bao}\ \emph {et~al.}(2023)\citenamefont {Bao}, \citenamefont {Fu}, \citenamefont {Pramanik}, \citenamefont {Mao}, \citenamefont {Chi}, \citenamefont {Cao}, \citenamefont {Zhai}, \citenamefont {Mao}, \citenamefont {Dai}, \citenamefont {Chen} \emph {et~al.}}]{bao2023very}%
  \BibitemOpen
  \bibfield  {author} {\bibinfo {author} {\bibfnamefont {J.}~\bibnamefont {Bao}}, \bibinfo {author} {\bibfnamefont {Z.}~\bibnamefont {Fu}}, \bibinfo {author} {\bibfnamefont {T.}~\bibnamefont {Pramanik}}, \bibinfo {author} {\bibfnamefont {J.}~\bibnamefont {Mao}}, \bibinfo {author} {\bibfnamefont {Y.}~\bibnamefont {Chi}}, \bibinfo {author} {\bibfnamefont {Y.}~\bibnamefont {Cao}}, \bibinfo {author} {\bibfnamefont {C.}~\bibnamefont {Zhai}}, \bibinfo {author} {\bibfnamefont {Y.}~\bibnamefont {Mao}}, \bibinfo {author} {\bibfnamefont {T.}~\bibnamefont {Dai}}, \bibinfo {author} {\bibfnamefont {X.}~\bibnamefont {Chen}},  \emph {et~al.},\ }\href@noop {} {\bibfield  {journal} {\bibinfo  {journal} {Nature Photonics}\ }\textbf {\bibinfo {volume} {17}},\ \bibinfo {pages} {573} (\bibinfo {year} {2023})}\BibitemShut {NoStop}%
\bibitem [{\citenamefont {Shekhar}\ \emph {et~al.}(2024)\citenamefont {Shekhar}, \citenamefont {Bogaerts}, \citenamefont {Chrostowski}, \citenamefont {Bowers}, \citenamefont {Hochberg}, \citenamefont {Soref},\ and\ \citenamefont {Shastri}}]{shekhar2024roadmapping}%
  \BibitemOpen
  \bibfield  {author} {\bibinfo {author} {\bibfnamefont {S.}~\bibnamefont {Shekhar}}, \bibinfo {author} {\bibfnamefont {W.}~\bibnamefont {Bogaerts}}, \bibinfo {author} {\bibfnamefont {L.}~\bibnamefont {Chrostowski}}, \bibinfo {author} {\bibfnamefont {J.~E.}\ \bibnamefont {Bowers}}, \bibinfo {author} {\bibfnamefont {M.}~\bibnamefont {Hochberg}}, \bibinfo {author} {\bibfnamefont {R.}~\bibnamefont {Soref}}, \ and\ \bibinfo {author} {\bibfnamefont {B.~J.}\ \bibnamefont {Shastri}},\ }\href@noop {} {\bibfield  {journal} {\bibinfo  {journal} {Nature Communications}\ }\textbf {\bibinfo {volume} {15}},\ \bibinfo {pages} {751} (\bibinfo {year} {2024})}\BibitemShut {NoStop}%
\bibitem [{\citenamefont {Shu}\ \emph {et~al.}(2022)\citenamefont {Shu}, \citenamefont {Chang}, \citenamefont {Tao}, \citenamefont {Shen}, \citenamefont {Xie}, \citenamefont {Jin}, \citenamefont {Netherton}, \citenamefont {Tao}, \citenamefont {Zhang}, \citenamefont {Chen} \emph {et~al.}}]{shu2022microcomb}%
  \BibitemOpen
  \bibfield  {author} {\bibinfo {author} {\bibfnamefont {H.}~\bibnamefont {Shu}}, \bibinfo {author} {\bibfnamefont {L.}~\bibnamefont {Chang}}, \bibinfo {author} {\bibfnamefont {Y.}~\bibnamefont {Tao}}, \bibinfo {author} {\bibfnamefont {B.}~\bibnamefont {Shen}}, \bibinfo {author} {\bibfnamefont {W.}~\bibnamefont {Xie}}, \bibinfo {author} {\bibfnamefont {M.}~\bibnamefont {Jin}}, \bibinfo {author} {\bibfnamefont {A.}~\bibnamefont {Netherton}}, \bibinfo {author} {\bibfnamefont {Z.}~\bibnamefont {Tao}}, \bibinfo {author} {\bibfnamefont {X.}~\bibnamefont {Zhang}}, \bibinfo {author} {\bibfnamefont {R.}~\bibnamefont {Chen}},  \emph {et~al.},\ }\href@noop {} {\bibfield  {journal} {\bibinfo  {journal} {Nature}\ }\textbf {\bibinfo {volume} {605}},\ \bibinfo {pages} {457} (\bibinfo {year} {2022})}\BibitemShut {NoStop}%
\bibitem [{\citenamefont {Yang}\ \emph {et~al.}(2022)\citenamefont {Yang}, \citenamefont {Shirpurkar}, \citenamefont {White}, \citenamefont {Zang}, \citenamefont {Chang}, \citenamefont {Ashtiani}, \citenamefont {Guidry}, \citenamefont {Lukin}, \citenamefont {Pericherla}, \citenamefont {Yang} \emph {et~al.}}]{yang2022multi}%
  \BibitemOpen
  \bibfield  {author} {\bibinfo {author} {\bibfnamefont {K.~Y.}\ \bibnamefont {Yang}}, \bibinfo {author} {\bibfnamefont {C.}~\bibnamefont {Shirpurkar}}, \bibinfo {author} {\bibfnamefont {A.~D.}\ \bibnamefont {White}}, \bibinfo {author} {\bibfnamefont {J.}~\bibnamefont {Zang}}, \bibinfo {author} {\bibfnamefont {L.}~\bibnamefont {Chang}}, \bibinfo {author} {\bibfnamefont {F.}~\bibnamefont {Ashtiani}}, \bibinfo {author} {\bibfnamefont {M.~A.}\ \bibnamefont {Guidry}}, \bibinfo {author} {\bibfnamefont {D.~M.}\ \bibnamefont {Lukin}}, \bibinfo {author} {\bibfnamefont {S.~V.}\ \bibnamefont {Pericherla}}, \bibinfo {author} {\bibfnamefont {J.}~\bibnamefont {Yang}},  \emph {et~al.},\ }\href@noop {} {\bibfield  {journal} {\bibinfo  {journal} {Nature communications}\ }\textbf {\bibinfo {volume} {13}},\ \bibinfo {pages} {7862} (\bibinfo {year} {2022})}\BibitemShut {NoStop}%
\bibitem [{\citenamefont {Riemensberger}\ \emph {et~al.}(2022)\citenamefont {Riemensberger}, \citenamefont {Kuznetsov}, \citenamefont {Liu}, \citenamefont {He}, \citenamefont {Wang},\ and\ \citenamefont {Kippenberg}}]{riemensberger2022photonic}%
  \BibitemOpen
  \bibfield  {author} {\bibinfo {author} {\bibfnamefont {J.}~\bibnamefont {Riemensberger}}, \bibinfo {author} {\bibfnamefont {N.}~\bibnamefont {Kuznetsov}}, \bibinfo {author} {\bibfnamefont {J.}~\bibnamefont {Liu}}, \bibinfo {author} {\bibfnamefont {J.}~\bibnamefont {He}}, \bibinfo {author} {\bibfnamefont {R.~N.}\ \bibnamefont {Wang}}, \ and\ \bibinfo {author} {\bibfnamefont {T.~J.}\ \bibnamefont {Kippenberg}},\ }\href@noop {} {\bibfield  {journal} {\bibinfo  {journal} {Nature}\ }\textbf {\bibinfo {volume} {612}},\ \bibinfo {pages} {56} (\bibinfo {year} {2022})}\BibitemShut {NoStop}%
\bibitem [{\citenamefont {Zhang}\ \emph {et~al.}(2022)\citenamefont {Zhang}, \citenamefont {Kwon}, \citenamefont {Henriksson}, \citenamefont {Luo},\ and\ \citenamefont {Wu}}]{zhang2022large}%
  \BibitemOpen
  \bibfield  {author} {\bibinfo {author} {\bibfnamefont {X.}~\bibnamefont {Zhang}}, \bibinfo {author} {\bibfnamefont {K.}~\bibnamefont {Kwon}}, \bibinfo {author} {\bibfnamefont {J.}~\bibnamefont {Henriksson}}, \bibinfo {author} {\bibfnamefont {J.}~\bibnamefont {Luo}}, \ and\ \bibinfo {author} {\bibfnamefont {M.~C.}\ \bibnamefont {Wu}},\ }\href@noop {} {\bibfield  {journal} {\bibinfo  {journal} {Nature}\ }\textbf {\bibinfo {volume} {603}},\ \bibinfo {pages} {253} (\bibinfo {year} {2022})}\BibitemShut {NoStop}%
\bibitem [{\citenamefont {Xu}\ \emph {et~al.}(2022{\natexlab{a}})\citenamefont {Xu}, \citenamefont {Guo}, \citenamefont {Li}, \citenamefont {Liu}, \citenamefont {Lu}, \citenamefont {Chen},\ and\ \citenamefont {Zhou}}]{xu2022fully}%
  \BibitemOpen
  \bibfield  {author} {\bibinfo {author} {\bibfnamefont {W.}~\bibnamefont {Xu}}, \bibinfo {author} {\bibfnamefont {Y.}~\bibnamefont {Guo}}, \bibinfo {author} {\bibfnamefont {X.}~\bibnamefont {Li}}, \bibinfo {author} {\bibfnamefont {C.}~\bibnamefont {Liu}}, \bibinfo {author} {\bibfnamefont {L.}~\bibnamefont {Lu}}, \bibinfo {author} {\bibfnamefont {J.}~\bibnamefont {Chen}}, \ and\ \bibinfo {author} {\bibfnamefont {L.}~\bibnamefont {Zhou}},\ }\href@noop {} {\bibfield  {journal} {\bibinfo  {journal} {Journal of Lightwave Technology}\ }\textbf {\bibinfo {volume} {41}},\ \bibinfo {pages} {832} (\bibinfo {year} {2022}{\natexlab{a}})}\BibitemShut {NoStop}%
\bibitem [{\citenamefont {Shen}\ \emph {et~al.}(2017)\citenamefont {Shen}, \citenamefont {Harris}, \citenamefont {Skirlo}, \citenamefont {Prabhu}, \citenamefont {Baehr-Jones}, \citenamefont {Hochberg}, \citenamefont {Sun}, \citenamefont {Zhao}, \citenamefont {Larochelle}, \citenamefont {Englund} \emph {et~al.}}]{shen2017deep}%
  \BibitemOpen
  \bibfield  {author} {\bibinfo {author} {\bibfnamefont {Y.}~\bibnamefont {Shen}}, \bibinfo {author} {\bibfnamefont {N.~C.}\ \bibnamefont {Harris}}, \bibinfo {author} {\bibfnamefont {S.}~\bibnamefont {Skirlo}}, \bibinfo {author} {\bibfnamefont {M.}~\bibnamefont {Prabhu}}, \bibinfo {author} {\bibfnamefont {T.}~\bibnamefont {Baehr-Jones}}, \bibinfo {author} {\bibfnamefont {M.}~\bibnamefont {Hochberg}}, \bibinfo {author} {\bibfnamefont {X.}~\bibnamefont {Sun}}, \bibinfo {author} {\bibfnamefont {S.}~\bibnamefont {Zhao}}, \bibinfo {author} {\bibfnamefont {H.}~\bibnamefont {Larochelle}}, \bibinfo {author} {\bibfnamefont {D.}~\bibnamefont {Englund}},  \emph {et~al.},\ }\href@noop {} {\bibfield  {journal} {\bibinfo  {journal} {Nature photonics}\ }\textbf {\bibinfo {volume} {11}},\ \bibinfo {pages} {441} (\bibinfo {year} {2017})}\BibitemShut {NoStop}%
\bibitem [{\citenamefont {Filipovich}\ \emph {et~al.}(2022)\citenamefont {Filipovich}, \citenamefont {Guo}, \citenamefont {Al-Qadasi}, \citenamefont {Marquez}, \citenamefont {Morison}, \citenamefont {Sorger}, \citenamefont {Prucnal}, \citenamefont {Shekhar},\ and\ \citenamefont {Shastri}}]{filipovich2022silicon}%
  \BibitemOpen
  \bibfield  {author} {\bibinfo {author} {\bibfnamefont {M.~J.}\ \bibnamefont {Filipovich}}, \bibinfo {author} {\bibfnamefont {Z.}~\bibnamefont {Guo}}, \bibinfo {author} {\bibfnamefont {M.}~\bibnamefont {Al-Qadasi}}, \bibinfo {author} {\bibfnamefont {B.~A.}\ \bibnamefont {Marquez}}, \bibinfo {author} {\bibfnamefont {H.~D.}\ \bibnamefont {Morison}}, \bibinfo {author} {\bibfnamefont {V.~J.}\ \bibnamefont {Sorger}}, \bibinfo {author} {\bibfnamefont {P.~R.}\ \bibnamefont {Prucnal}}, \bibinfo {author} {\bibfnamefont {S.}~\bibnamefont {Shekhar}}, \ and\ \bibinfo {author} {\bibfnamefont {B.~J.}\ \bibnamefont {Shastri}},\ }\href@noop {} {\bibfield  {journal} {\bibinfo  {journal} {Optica}\ }\textbf {\bibinfo {volume} {9}},\ \bibinfo {pages} {1323} (\bibinfo {year} {2022})}\BibitemShut {NoStop}%
\bibitem [{\citenamefont {Xu}\ \emph {et~al.}(2022{\natexlab{b}})\citenamefont {Xu}, \citenamefont {Wang}, \citenamefont {Yi},\ and\ \citenamefont {Zou}}]{xu2022high}%
  \BibitemOpen
  \bibfield  {author} {\bibinfo {author} {\bibfnamefont {S.}~\bibnamefont {Xu}}, \bibinfo {author} {\bibfnamefont {J.}~\bibnamefont {Wang}}, \bibinfo {author} {\bibfnamefont {S.}~\bibnamefont {Yi}}, \ and\ \bibinfo {author} {\bibfnamefont {W.}~\bibnamefont {Zou}},\ }\href@noop {} {\bibfield  {journal} {\bibinfo  {journal} {Nature communications}\ }\textbf {\bibinfo {volume} {13}},\ \bibinfo {pages} {7970} (\bibinfo {year} {2022}{\natexlab{b}})}\BibitemShut {NoStop}%
\bibitem [{\citenamefont {Hua}\ \emph {et~al.}(2025)\citenamefont {Hua}, \citenamefont {Divita}, \citenamefont {Yu}, \citenamefont {Peng}, \citenamefont {Roques-Carmes}, \citenamefont {Su}, \citenamefont {Chen}, \citenamefont {Bai}, \citenamefont {Zou}, \citenamefont {Zhu} \emph {et~al.}}]{hua2025integrated}%
  \BibitemOpen
  \bibfield  {author} {\bibinfo {author} {\bibfnamefont {S.}~\bibnamefont {Hua}}, \bibinfo {author} {\bibfnamefont {E.}~\bibnamefont {Divita}}, \bibinfo {author} {\bibfnamefont {S.}~\bibnamefont {Yu}}, \bibinfo {author} {\bibfnamefont {B.}~\bibnamefont {Peng}}, \bibinfo {author} {\bibfnamefont {C.}~\bibnamefont {Roques-Carmes}}, \bibinfo {author} {\bibfnamefont {Z.}~\bibnamefont {Su}}, \bibinfo {author} {\bibfnamefont {Z.}~\bibnamefont {Chen}}, \bibinfo {author} {\bibfnamefont {Y.}~\bibnamefont {Bai}}, \bibinfo {author} {\bibfnamefont {J.}~\bibnamefont {Zou}}, \bibinfo {author} {\bibfnamefont {Y.}~\bibnamefont {Zhu}},  \emph {et~al.},\ }\href@noop {} {\bibfield  {journal} {\bibinfo  {journal} {Nature}\ }\textbf {\bibinfo {volume} {640}},\ \bibinfo {pages} {361} (\bibinfo {year} {2025})}\BibitemShut {NoStop}%
\bibitem [{\citenamefont {Feldmann}\ \emph {et~al.}(2021)\citenamefont {Feldmann}, \citenamefont {Youngblood}, \citenamefont {Karpov}, \citenamefont {Gehring}, \citenamefont {Li}, \citenamefont {Stappers}, \citenamefont {Le~Gallo}, \citenamefont {Fu}, \citenamefont {Lukashchuk}, \citenamefont {Raja} \emph {et~al.}}]{feldmann2021parallel}%
  \BibitemOpen
  \bibfield  {author} {\bibinfo {author} {\bibfnamefont {J.}~\bibnamefont {Feldmann}}, \bibinfo {author} {\bibfnamefont {N.}~\bibnamefont {Youngblood}}, \bibinfo {author} {\bibfnamefont {M.}~\bibnamefont {Karpov}}, \bibinfo {author} {\bibfnamefont {H.}~\bibnamefont {Gehring}}, \bibinfo {author} {\bibfnamefont {X.}~\bibnamefont {Li}}, \bibinfo {author} {\bibfnamefont {M.}~\bibnamefont {Stappers}}, \bibinfo {author} {\bibfnamefont {M.}~\bibnamefont {Le~Gallo}}, \bibinfo {author} {\bibfnamefont {X.}~\bibnamefont {Fu}}, \bibinfo {author} {\bibfnamefont {A.}~\bibnamefont {Lukashchuk}}, \bibinfo {author} {\bibfnamefont {A.~S.}\ \bibnamefont {Raja}},  \emph {et~al.},\ }\href@noop {} {\bibfield  {journal} {\bibinfo  {journal} {Nature}\ }\textbf {\bibinfo {volume} {589}},\ \bibinfo {pages} {52} (\bibinfo {year} {2021})}\BibitemShut {NoStop}%
\bibitem [{\citenamefont {Galiffi}\ \emph {et~al.}(2022)\citenamefont {Galiffi}, \citenamefont {Tirole}, \citenamefont {Yin}, \citenamefont {Li}, \citenamefont {Vezzoli}, \citenamefont {Huidobro}, \citenamefont {Silveirinha}, \citenamefont {Sapienza}, \citenamefont {Al{\`u}},\ and\ \citenamefont {Pendry}}]{galiffi2022photonics}%
  \BibitemOpen
  \bibfield  {author} {\bibinfo {author} {\bibfnamefont {E.}~\bibnamefont {Galiffi}}, \bibinfo {author} {\bibfnamefont {R.}~\bibnamefont {Tirole}}, \bibinfo {author} {\bibfnamefont {S.}~\bibnamefont {Yin}}, \bibinfo {author} {\bibfnamefont {H.}~\bibnamefont {Li}}, \bibinfo {author} {\bibfnamefont {S.}~\bibnamefont {Vezzoli}}, \bibinfo {author} {\bibfnamefont {P.~A.}\ \bibnamefont {Huidobro}}, \bibinfo {author} {\bibfnamefont {M.~G.}\ \bibnamefont {Silveirinha}}, \bibinfo {author} {\bibfnamefont {R.}~\bibnamefont {Sapienza}}, \bibinfo {author} {\bibfnamefont {A.}~\bibnamefont {Al{\`u}}}, \ and\ \bibinfo {author} {\bibfnamefont {J.~B.}\ \bibnamefont {Pendry}},\ }\href@noop {} {\bibfield  {journal} {\bibinfo  {journal} {Advanced Photonics}\ }\textbf {\bibinfo {volume} {4}},\ \bibinfo {pages} {014002} (\bibinfo {year} {2022})}\BibitemShut {NoStop}%
\bibitem [{\citenamefont {Wang}\ \emph {et~al.}(2019)\citenamefont {Wang}, \citenamefont {Qiu}, \citenamefont {Xiao}, \citenamefont {Zhan}, \citenamefont {Bian}, \citenamefont {Yi},\ and\ \citenamefont {Xue}}]{wang2019simulating}%
  \BibitemOpen
  \bibfield  {author} {\bibinfo {author} {\bibfnamefont {K.}~\bibnamefont {Wang}}, \bibinfo {author} {\bibfnamefont {X.}~\bibnamefont {Qiu}}, \bibinfo {author} {\bibfnamefont {L.}~\bibnamefont {Xiao}}, \bibinfo {author} {\bibfnamefont {X.}~\bibnamefont {Zhan}}, \bibinfo {author} {\bibfnamefont {Z.}~\bibnamefont {Bian}}, \bibinfo {author} {\bibfnamefont {W.}~\bibnamefont {Yi}}, \ and\ \bibinfo {author} {\bibfnamefont {P.}~\bibnamefont {Xue}},\ }\href@noop {} {\bibfield  {journal} {\bibinfo  {journal} {Physical review letters}\ }\textbf {\bibinfo {volume} {122}},\ \bibinfo {pages} {020501} (\bibinfo {year} {2019})}\BibitemShut {NoStop}%
\bibitem [{\citenamefont {Rechtsman}\ \emph {et~al.}(2013)\citenamefont {Rechtsman}, \citenamefont {Zeuner}, \citenamefont {Plotnik}, \citenamefont {Lumer}, \citenamefont {Podolsky}, \citenamefont {Dreisow}, \citenamefont {Nolte}, \citenamefont {Segev},\ and\ \citenamefont {Szameit}}]{rechtsman2013photonic}%
  \BibitemOpen
  \bibfield  {author} {\bibinfo {author} {\bibfnamefont {M.~C.}\ \bibnamefont {Rechtsman}}, \bibinfo {author} {\bibfnamefont {J.~M.}\ \bibnamefont {Zeuner}}, \bibinfo {author} {\bibfnamefont {Y.}~\bibnamefont {Plotnik}}, \bibinfo {author} {\bibfnamefont {Y.}~\bibnamefont {Lumer}}, \bibinfo {author} {\bibfnamefont {D.}~\bibnamefont {Podolsky}}, \bibinfo {author} {\bibfnamefont {F.}~\bibnamefont {Dreisow}}, \bibinfo {author} {\bibfnamefont {S.}~\bibnamefont {Nolte}}, \bibinfo {author} {\bibfnamefont {M.}~\bibnamefont {Segev}}, \ and\ \bibinfo {author} {\bibfnamefont {A.}~\bibnamefont {Szameit}},\ }\href@noop {} {\bibfield  {journal} {\bibinfo  {journal} {Nature}\ }\textbf {\bibinfo {volume} {496}},\ \bibinfo {pages} {196} (\bibinfo {year} {2013})}\BibitemShut {NoStop}%
\bibitem [{\citenamefont {Maczewsky}\ \emph {et~al.}(2017)\citenamefont {Maczewsky}, \citenamefont {Zeuner}, \citenamefont {Nolte},\ and\ \citenamefont {Szameit}}]{maczewsky2017observation}%
  \BibitemOpen
  \bibfield  {author} {\bibinfo {author} {\bibfnamefont {L.~J.}\ \bibnamefont {Maczewsky}}, \bibinfo {author} {\bibfnamefont {J.~M.}\ \bibnamefont {Zeuner}}, \bibinfo {author} {\bibfnamefont {S.}~\bibnamefont {Nolte}}, \ and\ \bibinfo {author} {\bibfnamefont {A.}~\bibnamefont {Szameit}},\ }\href@noop {} {\bibfield  {journal} {\bibinfo  {journal} {Nature communications}\ }\textbf {\bibinfo {volume} {8}},\ \bibinfo {pages} {13756} (\bibinfo {year} {2017})}\BibitemShut {NoStop}%
\bibitem [{\citenamefont {Xu}\ \emph {et~al.}(2022{\natexlab{c}})\citenamefont {Xu}, \citenamefont {Gao}, \citenamefont {Krishna}, \citenamefont {Steinhauer}, \citenamefont {Zwiller},\ and\ \citenamefont {Elshaari}}]{xu2022direct}%
  \BibitemOpen
  \bibfield  {author} {\bibinfo {author} {\bibfnamefont {Z.-S.}\ \bibnamefont {Xu}}, \bibinfo {author} {\bibfnamefont {J.}~\bibnamefont {Gao}}, \bibinfo {author} {\bibfnamefont {G.}~\bibnamefont {Krishna}}, \bibinfo {author} {\bibfnamefont {S.}~\bibnamefont {Steinhauer}}, \bibinfo {author} {\bibfnamefont {V.}~\bibnamefont {Zwiller}}, \ and\ \bibinfo {author} {\bibfnamefont {A.~W.}\ \bibnamefont {Elshaari}},\ }\href@noop {} {\bibfield  {journal} {\bibinfo  {journal} {Photonics Research}\ }\textbf {\bibinfo {volume} {10}},\ \bibinfo {pages} {2901} (\bibinfo {year} {2022}{\natexlab{c}})}\BibitemShut {NoStop}%
\bibitem [{\citenamefont {J{\"u}rgensen}\ \emph {et~al.}(2021)\citenamefont {J{\"u}rgensen}, \citenamefont {Mukherjee},\ and\ \citenamefont {Rechtsman}}]{jurgensen2021quantized}%
  \BibitemOpen
  \bibfield  {author} {\bibinfo {author} {\bibfnamefont {M.}~\bibnamefont {J{\"u}rgensen}}, \bibinfo {author} {\bibfnamefont {S.}~\bibnamefont {Mukherjee}}, \ and\ \bibinfo {author} {\bibfnamefont {M.~C.}\ \bibnamefont {Rechtsman}},\ }\href@noop {} {\bibfield  {journal} {\bibinfo  {journal} {Nature}\ }\textbf {\bibinfo {volume} {596}},\ \bibinfo {pages} {63} (\bibinfo {year} {2021})}\BibitemShut {NoStop}%
\bibitem [{\citenamefont {Cerjan}\ \emph {et~al.}(2020)\citenamefont {Cerjan}, \citenamefont {Wang}, \citenamefont {Huang}, \citenamefont {Chen},\ and\ \citenamefont {Rechtsman}}]{cerjan2020thouless}%
  \BibitemOpen
  \bibfield  {author} {\bibinfo {author} {\bibfnamefont {A.}~\bibnamefont {Cerjan}}, \bibinfo {author} {\bibfnamefont {M.}~\bibnamefont {Wang}}, \bibinfo {author} {\bibfnamefont {S.}~\bibnamefont {Huang}}, \bibinfo {author} {\bibfnamefont {K.~P.}\ \bibnamefont {Chen}}, \ and\ \bibinfo {author} {\bibfnamefont {M.~C.}\ \bibnamefont {Rechtsman}},\ }\href@noop {} {\bibfield  {journal} {\bibinfo  {journal} {Light: Science \& Applications}\ }\textbf {\bibinfo {volume} {9}},\ \bibinfo {pages} {178} (\bibinfo {year} {2020})}\BibitemShut {NoStop}%
\end{thebibliography}%

\end{document}